# Complexity-Efficient Enumeration Techniques for Soft-Input, Soft-Output Sphere Decoding

Konstantinos Nikitopoulos, *Member IEEE,* Dan Zhang, I-Wei Lai, and Gerd Ascheid, *Senior IEEE*

The authors are with the Institute for Integrated Signal Processing Systems,
RWTH Aachen University, D-52074, Aachen, Germany
(e-mail: Konstantinos.Nikitopoulos@iss.rwth-aachen.de)

This work has been supported by the UMIC Research Centre, RWTH Aachen University.

*Abstract* – In this paper two complexity efficient soft sphere-decoder modifications are proposed for computing the max-log LLR values in iterative MIMO systems, which avoid the costly, typically needed, full enumeration and sorting (FES) procedure during the tree traversal without compromising the max-log performance. It is shown that despite the resulting increase in the number of expanded nodes, they can be more computationally efficient than the typical soft sphere decoders by avoiding the unnecessary complexity of FES.

## I. INTRODUCTION

Iterative receiver processing for multiple-input, multiple-output (MIMO) transmission has been proposed as a very efficient way to achieve near-capacity and high-rate transmissions [1]. However, the highly increased computational intensity of the iterative methods prevents current practical implementations from meeting the theoretical performance limits while also achieving the actual latency/area/energy consumption requirements.

At the heart of the corresponding methods lies the well known sphere decoding (SD) [2,3] process. It is employed by the soft MIMO detector to calculate, by means of tree searching, the *a-posteriori* soft information of the received bits per iteration. In this work, the soft-input, soft-output SD (SISO-SD) targets the (exact) calculation of the max-log log-likelihood ratio values and it is performed once per iteration, in contrast to the list SD approaches [1] where the tree search takes place only at the first iteration but the list size cannot be optimized easily to minimize the complexity without compromising the bit error rate (BER) performance.

In order to efficiently (in terms of number of visited nodes) perform the SD tree traversal the Schnorr and Euchner enumeration is typically employed [4], according to which the nodes are visited in ascending order of their partial distance (PD) metrics. However, minimizing the number of visited or expanded nodes does not necessarily minimize the overall SISO-SD complexity. The additional computational effort required for reducing the number of visited nodes may significantly reduce, or even eliminate, the gains originating from the visited nodes reduction. Therefore, the number of visited or expanded nodes is not a sufficient complexity measure when comparing different enumeration methods since the computational effort devoted to their minimization is ignored.

The significance of the additional computational effort allocated to the visited nodes reduction is revealed in the context of non-iterative, soft-output SD (SO-SD). Schnorr-Euchner enumeration can be realized by calculating and sorting the PDs of all children nodes of an expanded parent one. This full enumeration and sorting (FES) is rather inefficient since it imposes the PD calculation of all children nodes, even if only a few of them will be actually visited. Additionally, it requires a hardware expensive and energy consuming sorting over all children nodes. In order to avoid FES but still preserve the corresponding reduction on the number of expanded nodes, several less computationally intensive, approaches have been proposed [5,6]. They split the given QAM constellation into sub-constellations for which the search order can be efficiently calculated by employing simple geometrical properties. Such proposed methods split the high order QAM constellations into PSK-like [5] or PAM-like [6] sub-constellations. Then, the final visiting order is calculated by exhaustive search over the constellation subsets, resulting in a reduced number of redundant calculations, as well as a reduced sorting order.

In the herein targeted SISO-SD case the presence of the *a-priori* information in the PD function prevents the applicability of such methods since the desirable geometrical properties are annihilated. Thus, employing the Schnorr-Euchner enumeration to reduce the number of expanded[1] nodes requires the implementation-costly FES. The proposed SISO-SD approaches avoid FES by allowing an increase on the number of expanded nodes compared to the Schnorr-Euchner case. They employ alternative search orders based either on the available *channel* or on the *a-priori* information. They also modify the radius constraint check, by introducing the *pruning metric* concept, for keeping the number of expanded nodes low and still preserve the max-log performance. It is shown that this combination of low number of expanded nodes and reduced enumeration effort results in significant complexity gains. The proposed channel-information-based SISO-SD can efficiently be applied during the initial iterative decoding iterations, while the prior-information-based strategy is shown to be more efficient over the following iterations. Finally, one of the strong aspects of the proposed modifications is that they allow an easy extension of the efficient implementation SO-SD [5] techniques to the SISO case.

## II. SOFT-INPUT SOFT-OUTPUT SPHERE DECODING

Within any MIMO channel utilization with $M_T$ transmit and $M_R$ receive antennas ($M_R \geq M_T$) the interleaved coded bits are grouped into blocks $B_t$ ($t = 1,...,M_T$), with the bipolar $k$-th bit $c_k \in \{+1,-1\}$ residing in block $B_{\lceil k/\log_2|S| \rceil}$, which are mapped onto the symbols $s_t$ of a constellation set $S$ of size $|S|$ by a given mapping function (e.g., Gray coding). The corresponding received $M_R \times 1$ vector $\mathbf{y}$ is, then, given by

$$\mathbf{y} = \mathbf{Hs} + \mathbf{n}, \tag{1}$$

---

[1] Note that we refer to "expanded" instead of "visited nodes" since, for typical SISO-SD, all calculation complexity (i.e., PD calculations) occurs when expanding a node (due to FES).

where $\mathbf{H}$ is the $M_R \times M_T$ complex channel matrix, $\mathbf{s} = [s_1, s_2, ..., s_{M_T}]^T$ is the transmitted symbol vector, and $\mathbf{n}$ is the noise vector, consisting of complex, zero-mean, Gaussian, i.i.d. samples with variance $2\sigma_n^2$. The channel matrix $\mathbf{H}$ can be QR decomposed as $\mathbf{H} = \mathbf{QR}$, with $\mathbf{Q}$ a unitary $M_R \times M_T$ matrix and $\mathbf{R}$ an $M_T \times M_T$ upper triangular matrix with elements $R_{i,j}$ and real-valued positive diagonal entries.

Since, in principle, implementation of the optimal ML detector is not feasible iterative methods have been proposed which exchange soft information between the MIMO detector and the channel decoder and achieve near–capacity performance [1]. This soft information is typically expressed in terms of log-likelihood ratios (LLRs). Under the max-log approximation and under the assumptions of statistically independent interleaved bits and spatially transmitted symbols the MIMO detector computes the LLR value of $c_k$ as [1,5]

$$L(c_k) \approx \min_{\mathbf{s} \in S_k^{-1}} \{d(\mathbf{s})\} - \min_{\mathbf{s} \in S_k^{+1}} \{d(\mathbf{s})\} \qquad (2)$$

where $d(\mathbf{s}) = d_{channel}(\mathbf{s}) + d_{prior}(\mathbf{s})$, with

$$d_{channel}(\mathbf{s}) = \frac{1}{2\sigma_n^2} \sum_{i=1}^{M_T} \left| y'_i - \sum_{j=i}^{M_T} R_{i,j} s_j \right|^2 \qquad (3)$$

being the *channel-based* part of the soft information, and

$$d_{prior}(\mathbf{s}) = -\sum_{i=1}^{M_T} \ln P[s_i] \qquad (4)$$

being the *a-priori* part of the soft information, which is always non negative. Additionally, $\mathbf{y}' = \mathbf{Q}^H \mathbf{y} = [y'_1, y'_2, ..., y'_{M_R}]^T$ and $S_k^{\pm 1}$ are the sub-sets of $S$ having their $k$-th bit value equal to $\pm 1$.

Eq. (2) shows how the max-log LLR calculation can be reformulated into two minimization problems (i.e., hard output SDs) over the different symbol subsets (i.e., $S_k^{\pm 1}$), per decoded bit.

For each minimization problem the corresponding tree has its root at level $i = M_T + 1$ and its leafs at level $i = 1$. The $d(\mathbf{s})$ values for any leaf can be calculated recursively by

$$D(\mathbf{s}_i) = D(\mathbf{s}_{i+1}) + \Delta(\mathbf{s}_i) = D(\mathbf{s}_{i+1}) + \Delta_{channel}(\mathbf{s}_i) + \Delta_{prior}(\mathbf{s}_i) \tag{5}$$

where $\mathbf{s}_i = [s_i, s_{i+1}, ..., s_{M_T}]^T$ are partial symbol vectors, $D(\mathbf{s}_{M_T+1}) = 0$,

$$\Delta_{channel}(\mathbf{s}_i) = \frac{1}{2\sigma_n^2}\left|y'_i - \sum_{j=i}^{M_T} R_{i,j} s_j\right|^2 \tag{6}$$

and

$$\Delta_{prior}(\mathbf{s}_i) = -\ln P[s_i] \tag{7}$$

with $D(\mathbf{s}_i)$ being the PD of the $\mathbf{s}_i$ node. Then $d(\mathbf{s}) = D(\mathbf{s}_1)$.

### A. Typical SISO-SD (T-SISO-SD)

*Depth-first* tree traversal with Schnorr-Euchner enumeration and radius reduction is assumed [5]. The initial radius is assumed infinite and whenever a leaf is reached with its corresponding squared radius $r^2$ being smaller than $D(\mathbf{s}_1)$ the radius is updated to $D(\mathbf{s}_1)$. When a node is met a constraint check takes place. If its $D(\mathbf{s}_i) \geq r^2$ this node, its children, as well as its not yet visited siblings, are pruned. This constraint check procedure can be generalized by introducing the concept of the *pruning metric* $P_M$ which is defined as the metric to be compared to $r^2$. Then the constraint check is described as $P_M \geq r^2$ where for the typical SISO case $P_M^{(T)} = D(\mathbf{s}_i)$. In order to define the search order, namely the child node (of the parent node $\mathbf{s}_{i+1}$) to be visited next, FES over all possible $\Delta(\mathbf{s}_i)$ values (equal to $|S|$) is required. For the non-iterative SO-SD, $\Delta(\mathbf{s}_i)$ is equivalent to the distance metric $\Delta_{channel}(\mathbf{s}_i)$ (see (6)). Then, due to the absence of prior information, the costly FES can be avoided by employing the methods of [4] and [5] (see Section I), which reduce the number of unneeded PD calculations and the sorting computational complexity. However, as already discussed in Section I, in the soft-input case the presence of $\Delta_{prior}(\mathbf{s}_i)$ makes the aforementioned methods inapplicable.

*B. Channel-Information-Based SISO-SD (Ch-SISO-SD)*

FES can be avoided and the applicability of the abovementioned methods can be preserved if the nodes are still visited in ascending order of their $\Delta_{channel}(\mathbf{s}_i)$. Then, the pruning metric should be modified accordingly to preserve the max-log performance of the algorithm and still restrict the number of expanded nodes at the minimum possible complexity. Thus, we propose

$$P_M^{(Ch)}(\mathbf{s}_i) = D(\mathbf{s}_{i+1}) + \Delta_{channel}(\mathbf{s}_i) + \min_{s_i}\{\Delta_{prior}(\mathbf{s}_i)\}. \tag{8}$$

Replacing the exact prior value for the examined symbol with the minimum over the whole set of symbols results in $P_M^{(Ch)}(\mathbf{s}_i) \leq P_M^{(T)}(\mathbf{s}_i)$. Since the $r^2$ calculation is the same as in the T-SISO-SD case, the number of expanded nodes is increased. However, as shown in Section III, the total required computations are decreased due to the complexity reduction per expanded node. Additionally, instead of the full sorting per expanded node, $M_T$ minimization operations over the $|S|$ values of *a-priori* symbol information are only required. The Ch-SISO-SD is expected to work efficiently over the initial iterations, where the channel information is still dominant.

*C. Prior-Information-Based SISO-SD (Pr-SISO-SD)*

Whenever the calculated, incoming to the SD, bit reliability increases (this is typically observed over later iterations of a converging iterative scheme [7]) the search order can be efficiently based on the *a-priori* information. In this case the nodes can be visited in ascending order of their $\Delta_{prior}(\mathbf{s}_i)$ values. This still requires $M_T$ full sorting operations of the prior values, but only once per iteration. In this case, the corresponding pruning metric becomes

$$P_M^{(Pr)}(\mathbf{s}_i) = D(\mathbf{s}_{i+1}) + \min_{s_i}\{\Delta_{channel}(\mathbf{s}_i)\} + \Delta_{prior}(\mathbf{s}_i), \tag{9}$$

with $P_M^{(\text{Pr})}(\mathbf{s}_i) \leq P_M^{(T)}(\mathbf{s}_i)$. Since the symbol for which $\Delta_{channel}(\mathbf{s}_i)$ is minimized can be easily found geometrically, the additional complexity is, at most, one PD calculation per expanded node (i.e., when none of the visited symbols is the one minimizing $\Delta_{channel}(\mathbf{s}_i)$ which is already calculated for the pruning metric). However, the simulation results show that despite this additional complexity there are significant overall savings originating from avoiding FES.

*D. Implementation Issues in a Flexible Receiver Architecture*

In flexible receiver architectures, capable of performing trade-offs between the processing complexity and the provided BER performance, both SO (for the first iteration) and SISO-SDs shall be supported. Since no additional functionalities to the ones of the SO-SD (i.e., tree traversal, pruning process, ordering, sorting and PD calculations) the proposed approaches may share the same basic SD architecture without significant, additional implementation cost. However, as already discussed in Sections II.C and II.D, they are characterized by complexity gains originating from the reduction of the needed sorting operations. In addition to these gains, Section III depicts the reduced processing needs of the proposed approaches, for PD plus pruning metric calculations, using the number of complex multiplications as an indicative measure.

## III. SIMULATIONS

A 4x4 MIMO system is assumed operating over a spatially and temporally uncorrelated flat Rayleigh channel. The encoded bits are mapped onto 16-QAM via Gray coding. A systematic $(5/7)_8$ recursive convolutional code of rate ½ is employed as well as a random interleaver of 9216 information bits. The log-MAP BCJR algorithm has been employed for channel decoding. The operation $E_b/N_0$ is 8 dB and the achieved BER is about $6 \cdot 10^{-4}$ after 6 iterations. The PSK-like method proposed in [5] is employed for calculating the search order of Ch-SISO-SD. In order to avoid redundant calculations when employing multiple SDs the Single-Tree-Search approach of [4] is applied after being modified to perform the (pruning) constraint checks based on the pruning metric instead of the exact calculated PD value. Fig. 1 shows that despite the increased average number of expanded nodes per MIMO channel utilization (dashed curves) the proposed approaches need, on the average, less multiplications for PD processing and pruning metric calculations (solid curves) than the T-SISO-SD in their range of efficiency. The Ch-SISO-SD is more complexity efficient over the initial iterations (where the channel information is dominant) and saves about 43% at iteration 2, while the Pr-SISO-SD is more efficient over subsequent iterations (where the bit reliabilities tend to increase) and saves about 69% at iteration 6. Similar results hold also for other $E_b / N_0$ values of interest with the Pr-SISO-SD, over the second iteration, to be less complex than the Ch-SISO-SD in higher SNRs and more complex in lower ones.

## IV. CONCLUSIONS

Two soft-input, soft-output SD methods have been proposed which avoid the typically required complexity-inefficient FES. The efficient operational range of the proposed approaches (in terms of iteration number) is complementary but both solutions, as well as the soft-output SD, share the common functionalities. This attribute makes them strong candidates for joint implementation on flexible architectures.

ACKNOWLEDGMENT

The authors would like to thank Ernst Martin Witte, Filippo Borlenghi, David Kammler and Chun-Hao Liao for their valuable support.

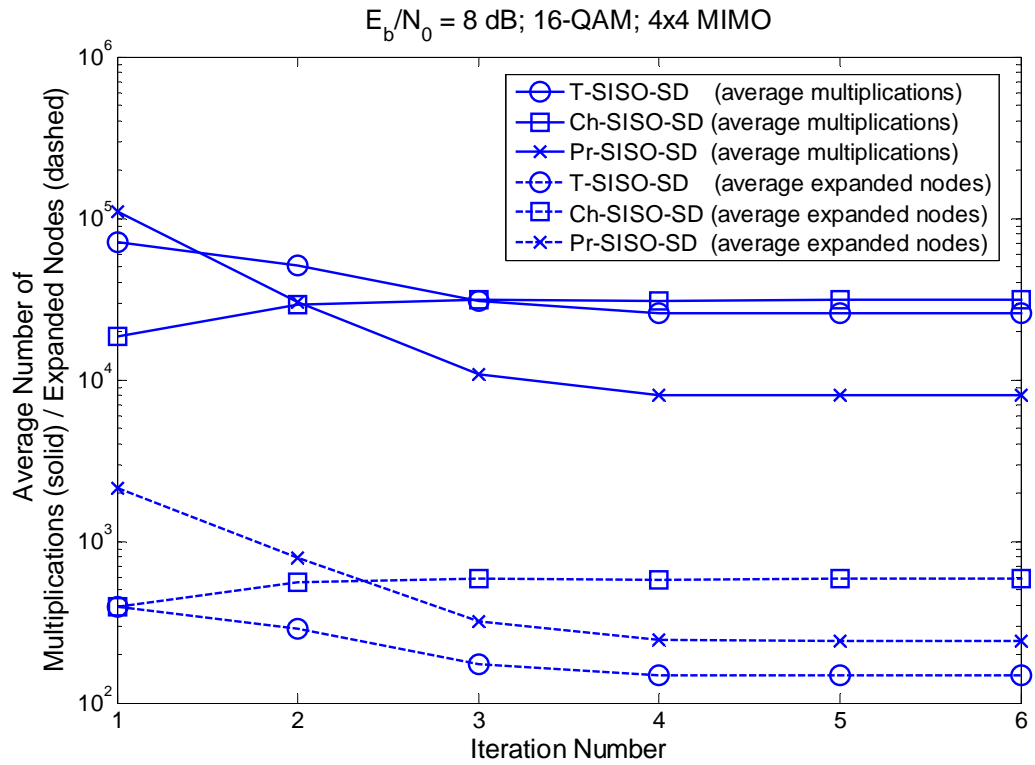

Fig. 1. Average (per MIMO channel utilization) number of expanded nodes and multiplications of the proposed methods over iterations.